\documentclass[preprint,psfig]{aastex}

\usepackage{graphicx}

\newcommand{\be}{\begin{equation}}
\newcommand{\ee}{\end{equation}}
\newcommand{\nn}{\mbox{} \nonumber \\ \mbox{} }
\newcommand{\ba}{\begin{eqnarray}}
\newcommand{\ea}{\end{eqnarray}}
\newcommand{\om}{\omega}
\newcommand{\Alfven}{ Alfv\'{e}n }
\newcommand{\curl}{{\rm curl\, }}
\newcommand{\E}{{\bf E}}
\newcommand{\B}{{\bf B}}
\newcommand{\J}{{\bf J}}
\renewcommand{\v}{{\bf v}}
\renewcommand{\k}{{\bf k}}
\renewcommand{\div}{{\rm \,div\,}}
\newcommand{\sech}{{\rm \,sech\,}}

\newcommand\etal{\textit{et al.\ }}
\newcommand\eg{\textit{e.g.\ }}

\newcommand\lo{\mathrel{\raise.3ex\hbox{$<$}\mkern-14mu\lower0.6ex\hbox{$\sim$}}}
\newcommand\go{\mathrel{\raise.3ex\hbox{$>$}\mkern-14mu\lower0.6ex\hbox{$\sim$}}}
\begin{document}
\date{}  
\title{Explosive reconnection in magnetars 
}
\author{MAXIM  LYUTIKOV$^{1,2}$\altaffilmark{3}  }
\affil{$^1$ Physics Department, McGill University, 3600 rue University
Montreal, QC,\\Canada H3A 2T8, \\
$^2$ Canadian Institute for Theoretical Astrophysics,\\ 60 St. George, Toronto, Ont,  
M5S 3H8, Canada}
\altaffiltext{3}{lyutikov@physics.mcgill.ca}

\begin{abstract}
X-ray activity of Anomalous X-ray
Pulsars and Soft Gamma-Ray Repeaters may result from the 
heating of their magnetic  corona
   by  direct currents dissipated by magnetic reconnection.
We investigate the  possibility that 
 X-ray flares and bursts observed from AXPs and SGRs 
result from magnetospheric 
 reconnection events initiated by 
 development of tearing mode in magnetically-dominated  relativistic  plasma.
 We  formulate equations of 
 resistive force-free 
 electrodynamics, discuss its relation to ideal  electrodynamics,
and give  examples of both ideal and  resistive equilibria.
Resistive  force-free current layers are 
unstable toward the development
of   small-scale  current sheets where 
resistive  effects become important. Thin current sheets are found to be
unstable due to the development of resistive force-free  tearing mode.
The growth rate 
 of    tearing mode
is  intermediate between the short \Alfven time scale
 $  \tau_A$ and a long resistive time scale $\tau_R$:
$\Gamma \sim 1/(\tau_R \tau_A)^{1/2}$,
 similar to the case of non-relativistic  non-force-free plasma.
 We propose that  growth of tearing mode is 
 related to the  typical
rise time of  flares,  $\sim 10$ msec. Finally, we discuss how reconnection
may explain other magnetar phenomena and ways to test the model.
\end{abstract}

\section{Introduction}

Two closely related classes of young neutron stars -- 
 Anomalous X-ray Pulsars (AXPs) and the Soft 
Gamma-ray Repeaters (SGRs) -- both show  X-ray flares and, once localized,
quiescent  X-ray emission (Kouveliotou et al. 1998, Gavriil \etal 2002; for
recent reviews see Mereghetti 2000, Thompson 2001). 
The energy powering the X-ray luminosity in these sources is 
supplied by the dissipation of  super-strong magnetic fields,
$B > 10^{15}$G (Thompson \& Duncan 1996). Hence the two classes are commonly referred to as magnetars.
  Magnetic fields are
 formed through a dynamo action during
 supernova collapse (Duncan \& Thompson 1992, Thompson \& Murray 2001).

A number of evidence suggest
 that processes that  lead to the production of X-ray flares
(and possibly of the persistent emission) on magnetars  is similar to those
operating in Solar corona (see also Section \ref{Discussion}; 
for alternative model addressing persistent  emission see Heyl \& Hernquist 1998).
The bursting 
activity of SGRs is strongly intermittent (good statistics for the AXPs'
bursting properties does not exists yet). 
The studies of  statistics of  SGR bursts from SGR 1900+14
(G\"og\"us et al. 1999)
 have found a power law  dependence of the number of flares on their energy, 
$dN/dE \sim E^{\alpha}$, with $\alpha=1.66$; this is   similar to  solar flares, 
where $\alpha=1.5-1.7$ 
                (Aschwanden \etal 2001).
The distribution of time intervals between successive
 bursts from SGR 1900+14 is 
 consistent with a log-normal distribution, also similar to
the
Sun (G\"og\"us et al. 1999). 
 In addition, Woods \etal (2001) have argued that the magnetic 
field of the neutron star
in SGR 1900+14 was
 significantly altered (perhaps globally) during the giant flare.
Given these similarities, 
 a suggestion that magnetar
bursts originate in the current-carrying magnetospheres is only natural.

Thompson,
 Lyutikov and
 Kulkarni (2002) have investigated the 
 global structure of   neutron star magnetospheres threaded
by large-scale electrical currents.
 In the magnetar
model for the Soft Gamma Repeaters and Anomalous X-ray Pulsars,
these currents are maintained by magnetic stresses acting
deep inside the star, which may  generate both sudden crustal fractures 
 and more gradual plastic deformations of the rigid
crust. 
They showed that  dissipation of the 
internal (to the neutron star) twisted magnetic field
may be 
 done efficiently in  much worse conducting magnetosphere where  currents
are pushed out by electro-magnetic torques.
In addition, a significant optical depth to  resonant cyclotron
 scattering is generated by the
current carriers. 
Resonant scattering in the magnetosphere
generates non-thermal
 component through Compton effect and modifies the pulse profile.

Several possible processes can lead to an explosive release of 
magnetic energy in  magnetosphere
(Thompson \& Duncan 1995,  Thompson \etal 2002).
 First, a sudden twist may be implanted into
the magnetosphere due to unwinding of the internal magnetic field.
This process is accompanied by a  large-scale displacement of the crust
 -- "gated" by the crust fracture properties.  This process is likely
to occur in a more brittle (as opposed to plastic) crusts.
Alternatively, in close analogy with Solar flares, 
a slow, plastic motion of the crust implants a twist (current) 
in the magnetosphere on a long time scale. At some point a  global
system of magnetospheric currents and sheared magnetic fields
 loses  equilibrium and produces
  a flare. This mechanism has the advantage 
that the energy stored in the external
twist need not be limited by the tensile strength of the crust, but instead
by the total external magnetic field energy. 
Both these mechanisms advertise that the magnetic  energy is released in the strongly
magnetized plasma through some kind of reconnection.
In this paper we investigate  the underlying  electro-magnetic 
processes responsible for this dissipation.
The two mechanisms  for the 
production of flares -- crust quake and loss of magnetic equilibrium -- 
are expected to have many similarities since they both involve 
dissipation of magnetic energy in the magnetosphere.
Thompson \etal (2002) and Lyutikov (2002)  have discussed  how these two
 possibilities
 may be distinguished 
observationally (see also Section \ref{Discussion}).

In our opinion the latter mechanism of flare production (loss of magnetic equilibrium)
is more promising.
Thus, we imagine that 
 the magnetosphere of magnetars are   similar to the solar corona: 
large multipoles contribute considerably to the total surface magnetic fields;
new current-carrying magnetic flux tubes rise  almost continuously 
 into the magnetosphere
gated by  rotational deformations of the neutron star crust; 
as a result,
magnetosphere consists of a complicated network of interacting current-carrying 
magnetic  flux tubes; coronal fields respond 
to slow injection of magnetic flux and currents by 
evolving through a series of quasi-static equilibria, 
which at some point   become unstable to resistive reconnection;
during relaxation some of the magnetic energy associated with the current is dissipated;
at a given time
magnetosphere may 
have several active regions, where the flux emergence is especially active;
dissipation of currents occurs on a wide variety of spatial and temporal scales in a very
intermittent fashion, with occasional giant flares which involve  the restructuring of the
whole magnetosphere.  
Overall, heating of the corona and production of flares
 is done by direct currents, which are dissipated by magnetic reconnection 
(\eg   Browning \& Priest 1986).

One of the arguments  raised against reconnection is that 
the typical time scales for the burst on-set, 
$\sim 10$ msec,
is much longer than
 the \Alfven travel time through magnetosphere, which is of the order
of the light travel time, $\sim 30 \, 
 \mu$sec (Thompson, private communication).
In this  paper we explore a possibility that the flare  on-set
is due to the development of a 
 tearing mode in the resistive magnetically-dominated magnetosphere
and show that the typical growth of  the tearing mode occurs on  a time-scale
intermediate between the very short \Alfven time scale and a long resistive time-scale.
 Tearing mode is 
one of the principle unstable  resistive  modes, which plays the main
role in various TOKAMAK discharges like sawtooth oscillations and  
major disruptions (\eg Kadomtsev 1975).  
Tearing mode is also the principle model for the unsteady
reconnection in Solar flares (\eg Shivamoggi 1985,  Aschwanden 2002) and Earth
magnetotail (\eg Galeev \etal 1978).

Reconnection in relativistic  magnetically-dominated plasma may be qualitatively
different  from the non-relativistic analogue. 
In force-free plasma   currents  are allowed to flow
only along  the magnetic  field in the plasma rest frame.
 It is not obvious that any reconnection may occur at all, since
in  conventional  models (\eg, Sweet-Parker), resistive currents flow mostly
across magnetic field.
One example of
 a qualitative difference of force-free and non-force-free fields
is a X-type point.
One cannot construct a non-trivial  (current-carrying)  X-type point
configuration of  a force-free field.
Another  important modification concerns Ohm's law.
In the force-free limit
  Ohm's law simplifies considerably since  both the inertia of both signs
of changes can be neglected and  there are
no Hall terms, which arise due to different masses of charge carries
\footnote{Hall term are also absent in pair plasma,
\eg Blackman and Field (1993)}.
This bears important consequences for the models of reconnection in 
relativistic magnetically 
dominated plasmas. In non-relativistic plasmas inertial terms in Ohm's law have
been invoked to produce what came to be known as collisionless reconnection.
Typically collisionless reconnection effects are related to ion inertial or 
cyclotron scales. In force-free plasmas (which  may be dominated
by pairs) no such effects arise. 

In addition, the 
conditions in 
magnetospheres of neutrons stars, and especially magnetars, are  
different from the better understood conditions in Solar
chromosphere in 
 several other  ways that are likely to influence reconnection.
 First,
 radiative cyclotron decays times are extremely short 
thus making particle distribution one-dimensional and forcing the currents
to flow exclusively along the field lines.
Secondly,
 the typical
collisional resistivities,
 which are far too small to explain the short dynamical time
scales of reconnection of the Sun, are even more suppressed by super-strong
magnetic fields (\eg for $k_B T \ll \hbar \om_B$ the suppression
is by  a factor  $\sim (k_B T / \hbar \om_B )^2$).

In spite of the obvious limitations of the 
resistive reconnection models,
 in particular formally long time scales, resistive 
 diffusion is known to be able to  drive instabilities
with time scales much shorter than the resistive
 time scale $\tau_R \sim L^2/\eta$.
This is done through a formation of a very narrow current sheet where both
the time scales for diffusion may be short and, in addition, resistivity may be 
enhances due to development of plasma turbulence (anomalous resistivity). 
The resulting current  
sheet tend to be unstable   to transverse, $\k \cdot \B=0$, perturbations.
This resistive  instability is called  a tearing mode.

\begin{figure}
\includegraphics[width=0.9\linewidth]{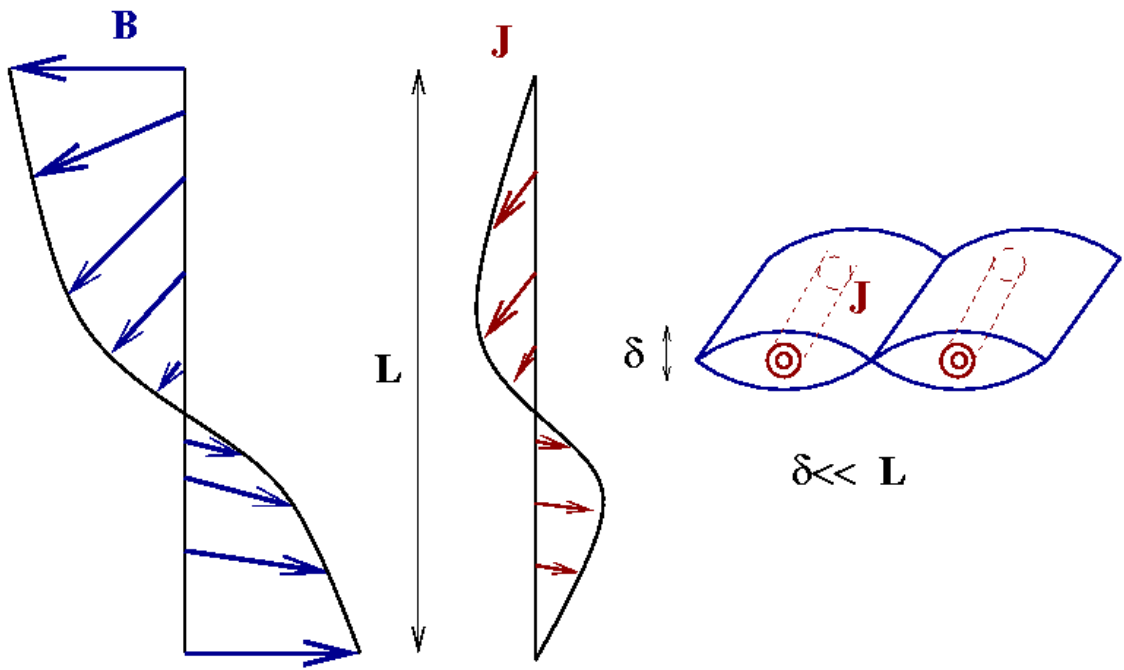}
\caption{ Development of a tearing mode. Initial configuration consists 
of a current layer in which  magnetic field rotates by $\pi$ radians. 
Current is flowing along magnetic field.
A thin current sheet forms due to 
 development of tearing mode. Magnetic islands 
form inside the resistive  current sheet.
 }
\label{tear-sheet}
\end{figure}

Tearing mode is quite
 complicated (Furth \etal 1963, White 1983, Fig. \ref{tear-sheet}). The basic
physical processes leading to development of tearing mode are the following.
A current sheet may be represented as a set of elementary current filaments.
Since current filaments attract each other the current sheet is unstable 
against pairing of current filaments. For sufficiently narrow current sheets the
energy released by pairing of currents is more than the energy spent to maintain 
perturbed magnetic field - tearing mode has negative energy, so that its dissipation 
should  increase its amplitude leading to dissipation. Magnetic islands form in plasma
as a result of the development of tearing mode.
Typical time scale for the development of the tearing mode is intermediate between a
short \Alfven and long resistive time scale.

\section{Resistive force-free electrodynamics}

Under force-free approximation it is assumed that  the plasma dynamics
is completely controlled by magnetic field. The validity of this approximation
requires $b^2 \gg \rho c^2$, where $b$ and $\rho$ are plasma rest-frame 
magnetic field and density. It is assumed that the plasma provides currents and charge
densities required by the dynamics of electro-magnetic fields, but these current carry no 
inertia.

In addition to being magnetically-dominated,
 microscopic plasma processes, like particle collisions or 
 plasma turbulence, may
   contribute to resistivity and thus  make plasma  non-ideal.
 Resistivity will result in the decay of currents supporting the magnetic field; this, in tern,
will influence the plasma dynamics. 
We wish to explore how the decay of inertialess currents
will affect the dynamics of the system. 
We assume that plasma resistivity  can be represented by 
phenomenological parameter $\eta$. 
To calculate $\eta$ from microscopical principles one needs to take the particle dynamics
into account and thus go beyond the inertialess approximation. The answer will depend 
on the types of current-carrying particle (\eg electron-ion or pair plasma) and on the
 dominant type of  scattering (\eg particle-particle, particle-wave or 
particle-wave-particle). We wish to avoid this complication by 
introducing a macroscopic parameter  $\eta$. 
We also 
 assume that the dissipated energy of the magnetic field
leaves the system, \eg in a form of radiation. If this was not the case, the particle
pressure would build up to equipartition breaking the force-free assumption.

Strictly speaking, inclusion of resistivity violates the force-free condition since  
non-zero resistance implies  that there is 
 extra force acting on charge carries. Still, in strongly magnetized
plasmas the resistive forces act only along the direction of the 
magnetic field, while the  dynamics  transverse to the magnetic field remains
unaffected, given by the force-free conditions (\ref{Ohm-2}). 
What distinguishes resistive and ideal cases is the way the current is related to the field
(Ohm's law). It is in this sense that  
  we will use the term resistive force-free electro-dynamics.

The procedure described above is not entirely consistent in case 
of relativistic plasmas. The reason is 
that in an ideal force-free plasma the velocity along the field is not
 defined. Since  plasma resistivity must be defined in the plasma rest-frame
this creates a principal ambiguity. Note, that in case of non-relativistic force-free
 plasma 
such ambiguity does not appear: it is possible to write a self-consistent
system of equations describing evolution of  non-relativistic
resistive force-free plasma (Chandrasekhar \& Kendall 1957,
Low 1973).
In the non-relativistic plasma evolution of resistive force-free fields
proceeds in a self-similar ways:  magnetic fields keep their configuration
while changing in magnitude. In  a process of doing so small electric
fields and charge densities develop, whose  dynamical contribution is neglected in the
 non-relativistic case.
 In relativistic case we cannot neglect these
effects so that  evolution  force-free resistive fields becomes much more complicated.

\subsection{Ohm's law in magnetically-dominated  plasma}

Relativistic form of Ohm's law in magnetized plasma may be formally written
assuming that in the plasma rest frame the current is
proportional to the  electric field. Several attempts have been made to derive 
Ohm's law in relativistic plasma; none seem completely satisfactory so far.
Blackman and Field (1993) assumed that there exists a frame where charge density,
momentum density and velocity flux density are all zero, but in a general case
 the plasma may be charged in its rest frame (defined, 
for example, as a frame where
 momentum density is zero). Gedalin (1996)  wrote expressly covariant form of the
relativistic Ohm's law in isotropic plasma. 
In strong magnetic field we expect that transverse
 and parallel (with respect to magnetic field)
conductivities are different,  so that conductivity becomes a tensor: 
\be
F^{i k} U_k = \eta^{ik} \left( \delta _k ^l - U_k U^l \right) j_l
\label{OhmM}
\ee
where 
$\eta^{ik}$ is a conductivity  tensor. 
To find an expression for $ \eta^{ik}$ we need to consider particle dynamics
taking into account particle-particle interaction.
Derivation of a general expression for $ \eta^{ik}$  is  a project worth a separate
paper. Here we give simplified expressions
 for $ \eta^{ik}$ in case of strongly magnetized plasma.

To derive Ohm's law 
in force-free plasma, which assumes a one-fluid description, 
we need to define the plasma rest-frame. In relativistic plasma the choice
of  plasma rest-frame is not unique (\eg de Groot \etal 1980). 
In a force-free plasma the ambiguity of the rest frame is partially removed: 
though the motion along the magnetic field still remains undefined, 
the motion of particles across the
magnetic field consists only of an electric drift
\be 
\v_\perp = { \E_\perp \times \B \over B^2}, 
\ee
which allows one to introduce a plasma frame (up to a boost along the field),
 as a frame where drift velocity vanishes.
Since both species drift with the same velocity, there is no relative motion
of particles across the field and thus there is no resistivity across the 
magnetic  field
 (the current across
the field is due to exclusively different charge densities of two species). 
An alternative way of reasoning supporting this conclusion is that in force-free
plasma  there is no current 
across the field in the plasma rest-frame, where resistivity should be defined.

Thus, under force-free approximation the  resistive effects affect
only the motion along the field. To derive the corresponding Ohm's law
we introduce a four-dimensional magnetic field vector
\be
b_i = \ast F_{i k} U^k
\ee
where $\ast F_{i k}$ is a dual electro-magnetic tensor. 
If $\B$ denotes the plasma rest frame magnetic field, then
$B_i=\{0, \B\}$  and 
\be 
b_i = \gamma \{ (\v \cdot \B), \B \}
\ee
Convolving eq. (\ref{OhmM}) with $b_i$ we find
\be
F^{i k} U_k  \ast F_{i l} U^l =
 \eta^{ik} b_i \left( \delta _k ^l - U_k U^l \right) j_l
\label{OhmM1}
\ee
Since we expect that in the strongly magnetized plasma resistivity is 
important only along the field, we assume
\be 
\eta ^{ik} = \eta {b^i b^k \over b^2}
\ee

We find then
\be
F^{i k} U_k  \ast F_{i l} U^l = \eta (j b) = \eta  \ast F_{i l} U^l j^i
\label{FFOhm1}
\ee
since $(bU)=0$. 
Equation (\ref{FFOhm1}) gives  Ohm's law in force-free plasma.
\footnote{Another possibility to define Lorentz-invariant   Ohm's law in force-free plasma (suggested by Blandford, private communication)
is  to postulate that $ \eta (j b) = F^{i k} \ast F_{ik} = \E \cdot \B$.
The two definitions are  
different by  the term involving $(\B \cdot \v) (\E\cdot \v)$ in (\ref{M}).}
In 3-D notations this gives
\be
\gamma^2 
\left( (\E\cdot \v) (\B \cdot \v) - 
( \E + \v \times \B ) \cdot (\B + \v \times \E)  \right) =
\eta \gamma \left(j_0 (\B \cdot \v) -(\J \cdot \B) \right)
\label{M}
\ee
Next we separate electric fields and velocity into components along and across
the magnetic field, $\E = \E _\parallel + \E_\perp $,
$\v =\v_\parallel +\v_\perp$ with $\v_\perp= \E_\perp \times \B /B^2$.
Then eq. (\ref{M}) simplifies
\be E _\parallel
\left(   v_\parallel ^2 -  1  \right) 
= {\eta \over \gamma} 
\left(j_0   v_\parallel - J_\parallel ) \right)
\label{M1}
\ee
Since the velocity along the field cannot be specified within the framework
of force-free plasma we chose $v_\parallel =0$. 
The ambiguity for the choice of $v_\parallel$
 stems from the fact that the force-free approximation
cannot in principle 
describe the plasma dynamics along the magnetic  fields -- however
strong the field 
is, it does not affect particle motion align the field, so that one
must  use 
full  MHD  equations. 
For example, in a frame-work of two-fluid hydrodynamics there are two equations
of motion describing the  dynamics of each species along the field.
  The difference  of these equations
 gives 
the generalized Ohm's law. The sum gives the equation of bulk 
 motion of plasma due to the influence of external fields.
Since we neglect the Poynting flux associated with
the bulk motion this equation is neglected. Then the choice of parallel velocity
becomes just a choice of coordinate system.

 In spite of limitations concerning the parallel dynamics of plasma
 there is a number of 
relevant problems where 
one can neglect the plasma dynamics along the field either
because the variation in that direction are small or because of a symmetry
of the system (\eg if magnetic field is directed
along a cyclic variable).

Setting $v_\parallel =0$ in (\ref{M1}) we find
\be
 (\E \cdot \B)  ={\eta \over \gamma} (\J \cdot \B)
\label{OhmM2}
\ee
The factor $ \gamma = 1/\sqrt{ 1-(\E\times\B/B^2)^2}$
 comes from the Lorenz transformation of the electric field
 to the plasma rest frame ($J_\parallel$ is invariant under Lorentz transformation 
along the direction orthogonal to magnetic field). 
Reinstating the transverse current, we find  Ohm's law in 
relativistic force-free electro-dynamics
\be
\J = 
 { \div \E (\E \times \B) \over B^2} + 
{  1 \over  \eta  \sqrt{1 - \left( { (\E \times \B) \over B^2} \right)^2 }} 
 {({ \E \cdot \B}) \B \over B^2}
\label{ffOhm}
\ee
 Taking a vector product with $\B$ in Ohm's law we find a cross-field
 dynamical equation
\be
\J \times \B + \div \E \left( \E - { (\E\cdot\B) \over B^2} \B \right)
\equiv \J \times \B + \E_\perp \div \E 
=0
\label{Ohm-2}
\ee
This generalizes the force-free condition for resistive plasma.

\subsection{Equations of resistive force-free electro-dynamics}

The 
resistive
 force-free electrodynamics may be derived from RMHD in the limit of vanishing
plasma inertia $w,\,p,\, \rho \rightarrow 0$ and by reinstating
electric field 
\be
\E= \E_\parallel+ \E_\perp= { (\E \cdot \B) \B \over B^2}
 - {\bf v} \times \B.
\ee
In a force-free formulation 
the plasma velocity is defined 
only  up to an arbitrary Lorenz boost along the magnetic 
field direction. The velocity across
the magnetic field is just the  electro-magnetic drift
 velocity ${\bf E}\times{\bf B}/B^2$.
The dynamical equations are  Maxwell equations  
 and Ohm's law (\ref{ffOhm})
\ba && 
\dot{\B} = -  \curl \E
\nn &&
\dot{\E} = \curl \B -  \J
\label{Max}
\ea
(we use a system of units with the speed of light set to unity; we also
incorporate the coefficient $ 4 \pi$ into definitions of currents and charge
densities).

Resistive force-free electrodynamics is more complicated than its
non-relativistic analogue. In the non-relativistic case currents can be expressed
from the second Maxwell equation and electric field can be eliminated from Ohm's
law. In relativistic case this is impossible - the best that we can do is to eliminate
the current from Ohm's law and solve Maxwell  equations for a
$\B$ and $\E$ (this is a system of 5 equations  since  the 
constraint $\div \B=0$ must be satisfied at all times). 
Resistive force-free electrodynamics is also more complicated than the
ideal one. In ideal force-free electrodynamics the condition $\E \cdot \B =0$ 
reduced a number of equations to 4 (Komissarov 2002, Lyutikov \& Blandford 2003).
In addition, in ideal case the second  electro-magnetic invariant is always
larger than 0, $B^2 - E^2 >0$, which implies  that  there
is a  reference frame where  electric  field is  equal to
0. In case of resistive electrodynamics $B^2 - E^2 $ may be smaller than zero, which
implies an electric field in the plasma rest frame.

The form of Ohm's law (\ref{ffOhm}) may not be appropriate for some 
applications: for example,
 the transition to the ideal case ($\eta \rightarrow 0$) 
leave us with no conditions on the current. 
To make the transitions from the resistive to the ideal case clearer
we  combine Maxwell equations by taking scalar products with $\B$ and 
$\E$
\be 
 (\J \cdot \B) = 
({\bf B}\cdot\nabla\times{\bf B}-{\bf E}\cdot
\nabla\times{\bf E}){\bf B} -
\partial_t\left(\E \cdot \B \right) 
\label{jB}
\ee
The  term ${\bf B}\cdot\nabla\times{\bf B}$  is called current helicity, 
the corresponding term involving electric fields does not have a
 name (it does not appear
in non-relativistic analysis).

Using the  Ohm's law  (\ref{ffOhm})
we can eliminate 
$\left(\E \cdot \B \right)$ from  (\ref{jB})
\be
(\J \cdot \B) + \partial_t\left( \eta \sqrt{ 1- \left( (\E \times \B)/B^2 \right)^2} 
(\J \cdot \B) \right) = 
({\bf B}\cdot\nabla\times{\bf B}-{\bf E}\cdot
\nabla\times{\bf E}){\bf B}
\label{Ohm-1}
\ee
which gives an alternative form of Ohm's law.
This form  has an advantage that, unlike eq.  (\ref{ffOhm}),
it clearly shows how to make a transition to the 
ideal case.

Given Ohm's law (\ref{ffOhm}) we can write down the basic  law of energy
and momentum 
conservation in resistive force-free electro-dynamics:
\ba &&
{1 \over 2} 
\partial_t( B^2 +E^2) = \div \E \times \B - 
{  1 \over  \eta  \sqrt{1 - \left( { (\E \times \B) \over B^2} \right)^2 }} 
 {({ \E \cdot \B})^2  \over B^2}
\nn &&
\partial_t( \E \times \B ) =
\curl\B \times \B - \curl\E \times \E - \J\times \B =
\nn &&
\curl\B \times \B - \curl\E \times \E + 
\div \E \left( \E - { (\E\cdot\B) \over B^2} \B \right)
\ea

In ideal plasma, $\eta =0$, Ohm's law gives $\E \cdot \B=0$. This condition plus
 the Maxwell equations
(\ref{Max}) are then sufficient to express the current in terms of fields, 
\be
{\bf J}={({\bf E}\times{\bf B})\nabla\cdot{\bf E}+
({\bf B}\cdot\nabla\times{\bf B}-{\bf E}\cdot
\nabla\times{\bf E}){\bf B}\over   B^2}
\label{FF}
\ee
while eq. (\ref{Ohm-2}) gives
\be
\J \times \B +  \E \div  \E =0
\label{dyn}
\ee
Equations (\ref{Max}) and (\ref{FF}) (or (\ref{dyn}))  are then the 
equations of ideal force-free electrodynamics
 (Uchida 1997, Gruzinov 1999, Komissarov 2002, 
 Lyutikov \& Blandford 2003).

\subsection{Applicability of force-free approximation}

Force-free electro-dynamics assumes that inertia of plasma is negligible. 
This approximation is bound to break down for very large effective plasma
 velocities, when electric field becomes too close in value to magnetic field, 
$E \rightarrow B$. The condition that inertia is negligible is equivalent to the
condition that the  effective plasma four-velocity,
 $u  \sim E/(B\sqrt{1-(E/B)^2})$, is smaller that the
\Alfven four-velocity in plasma. Assume that in the plasma rest frame the ratio 
of magnetic energy density to  plasma energy density   (which may include both
rest mass energy density and internal energy) is $\sigma$.
 In  a strongly magnetized 
plasma $\sigma \gg 1$ (in  a force-free plasma  $\sigma$ is assumed to be infinite). 
Then the \Alfven wave phase velocity and the corresponding Lorentz factor  are
 (\eg Kennel \& Coroniti 1984, 
Lyutikov \& Blandford 2003)
\be 
u_A^2 = {\sigma } 
\ee
This puts an upper limit on the value of electric fields consistent with
force-free approximation
\be 
{B^2 -E^2 \over E^2} \gg {1 \over \sigma}
\ee

\section{Relativistic   Current Sheets}

\subsection{ Ideal  Force-free  Current Sheets}
\label{ideal}

Since we are interested in the stability of weakly resistive configurations, we first
consider 
 steady state planar solution of the ideal force-free electrodynamics (eqns. 
 (\ref{Max}) and (\ref{FF})).
Assume that magnetic field lays in the $x-y$ plane, $B_z=0$, and  
that all quantities dependent  only on  $z$.
Then from the first Maxwell equation we find that $E_x,\, E_y \sim {\rm const}$.
The $x$ and $z$ components of the $\curl \B=\J$ equation then give
($y$ component is the same as $x$)
\ba && 
\partial_z \left( B_x^2+B_y^2 -E_z^2 \right) =0
\nn && 
\left(B_y E_x - B_x E_y \right) \partial_z E_z =0
\ea
In addition, for ideal plasma condition $\E \cdot \B=0$ requires
\be 
B_x E_x +B_y E_y =0
\ee

Below we discuss  several possible solutions of these equations.
The two main types of current sheet equilibrium are the current sheets (when the
electric field is zero and there is no motion of plasma) and shear layer, where
non-zero electric fields  
lead to plasma motion in the plane of  the current sheet 
or into it, creating magnetic and  velocity-sheared configurations. 

\begin{figure}
\includegraphics[width=0.9\linewidth]{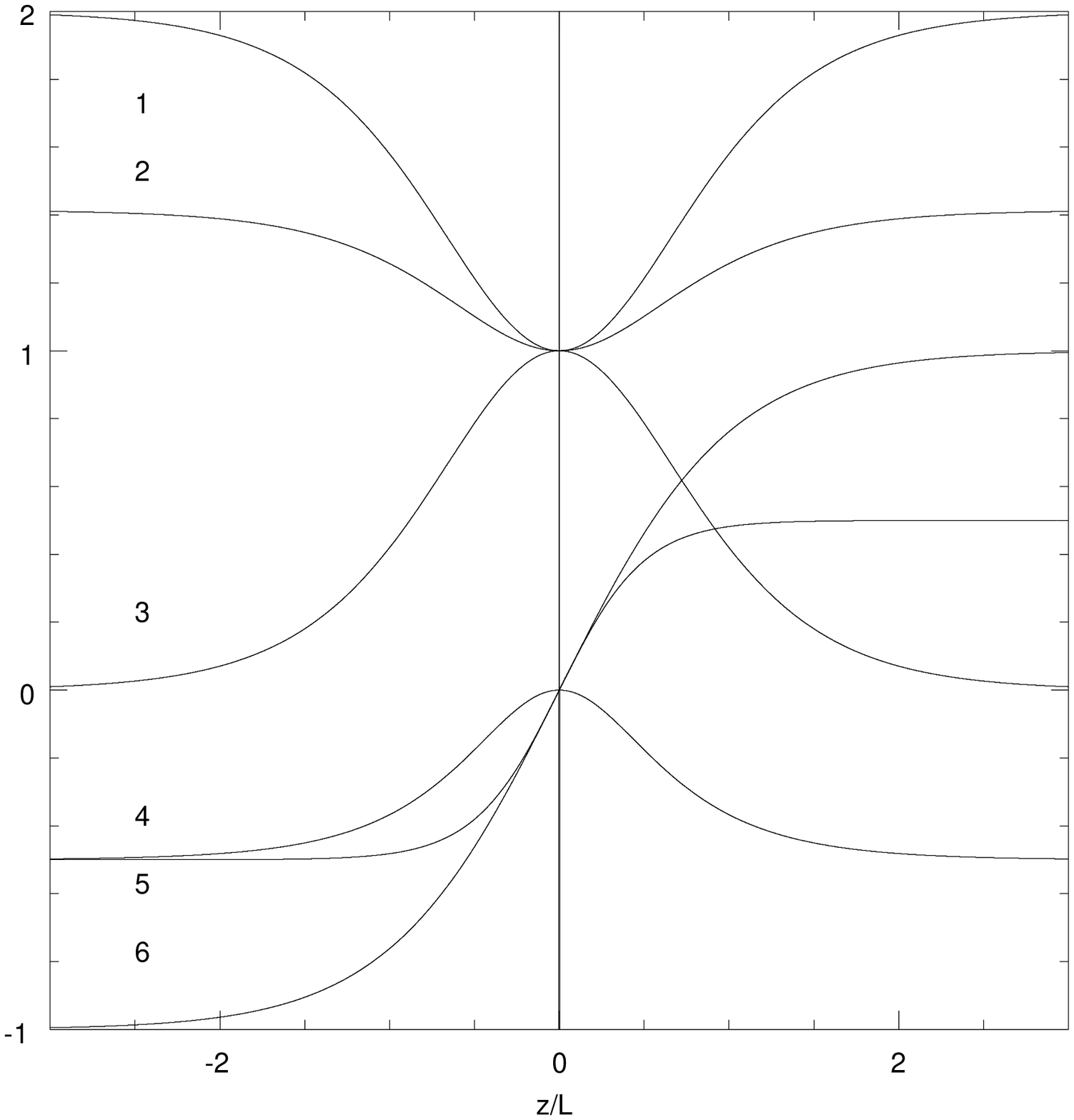} 
\caption{ First type of electro-magnetic sheared layer -- sheared $B_x$ field in a constant guiding field. 
At large $|z| \rightarrow \infty$
the medium is moving in opposite direction  with total  four-velocity $\pm U_0$;
 On the different sides of the
current layer the plasma is streaming in opposite direction along the $x$, with a
four velocity at infinity reaching $u_x = \pm U_0/\sqrt{1+U_0^2}$, while
along the $y$ direction the plasma is streaming in the same
direction on both sides of the layer, reaching at infinity 
$u_y = - U_0^2/\sqrt{1+U_0^2}$.
 The curves, plotted for $U_0=1$, 
correspond to (1) total Lorentz factor $\gamma$ , (2)  total magnetic
field $B^2 = B_x^2+B_y^2$, (3) current $j_y$, (4) velocity along  $y$ direction
$v_y$,
(5) velocity along  $x$ direction $v_x$,
(6) $B_x$.
  }
\label{shear-1}
\end{figure}

\begin{figure}
\includegraphics[width=0.9\linewidth]{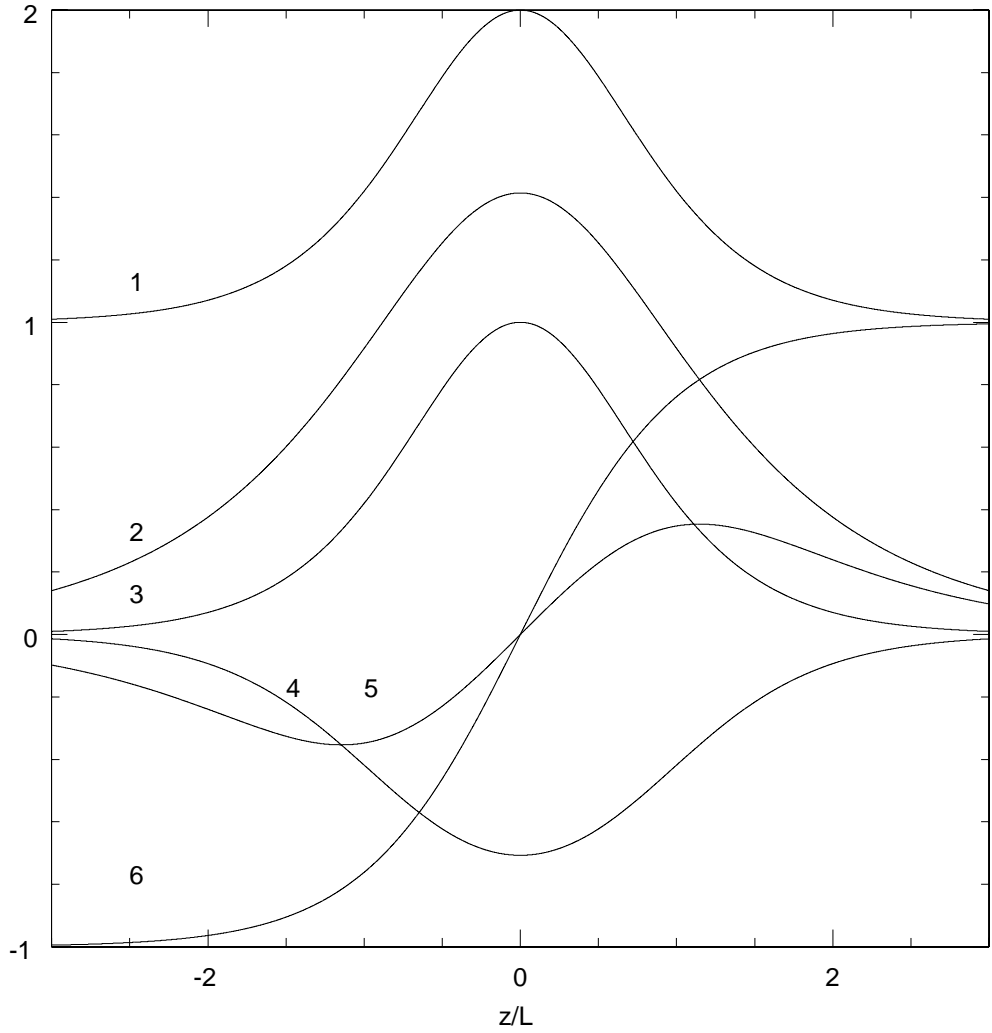}
\caption{Second type of  electro-magnetic sheared layer -- rotating and sheared
magnetic field.  The curves correspond to (1) total Lorentz factor $\gamma$ ,
(2) $B_y$, (3) current $j_y$, (4) velocity along  $x$  direction
$v_x$, (5) velocity along  $y$ direction
$v_y$, (6) $B_x$.
  }
\label{shear-2}
\end{figure}

\begin{enumerate}
\item {\it Crossed constant electric and magnetic fields}.
We assume that the  components of the electric field in the plane of the 
sheets  are non-zero, $E_x E_y \neq 0$ and, in addition,
 $E_z \sim {\rm const} $.
This case corresponds to plasma in constant electric and magnetic fields 
moving with velocities 
\be
{\bf V} =
\left\{ { E_z \over B_0 \sqrt {1+ E_y^2/E_x^2}}, 
-{E_y E_z \over B_0 E_x \sqrt {1+ E_y^2/E_x^2}},
{E_x^2 - E_y^2 \over  B_0 E_x \sqrt {1+ E_y^2/E_x^2}} \right\}
\ee
 (in Cartesian coordinates  $x,y,z$).
By Lorentz transformation electric fields can be completely eliminated, so that 
in the plasma rest frame $\E =0$.
\item
{\it Sheared force-free layer}. Assuming  $E_x =E_y =0$, $E_z \neq 0$
we find 
\be
B_x^2 +B_y^2 -E_z^2 = B_0^2
\label{EE}
\ee
where $B_0$ is some constant.
There are two interesting types of equilibrium here. 
First is a sheared $B_x$ field in a constant guiding field
 $B_y=  B_0/U_0$ ($U_0$ is  some
constant).
\ba &&
\B = B_0  \left\{ \tanh z/L, {1 \over U_0} ,0 \right\}
\nn &&
\E = B_0 \left\{ 0,0, \tanh z/L\right\}
\nn &&
\J = {B_0 \over L}  \left\{ 0 ,  \sech^2 z/L,0 \right\}
\nn &&
\rho_e = - {B_0 \over L} \sech z/L \tanh z/L 
\nn &&
{\bf V} =
\left\{   {  U_0  \tanh z/L  \over  1 +  U_0^2 \tanh ^2  z/L}, 
- { U_0^2  \tanh ^2  z/L \over 1 + U_0^2 \tanh ^2  z/L}  ,0\right\}
\nn &&
 v^2 = { U_0^2 \tanh ^2  z/L \over 1 +U_0^2  \tanh ^2  z/L}, \,
\gamma^2 = 1+ U_0^2 \tanh ^2  z/L
\nn &&
{\bf P}= B_0^2
\left\{ {1\over U_0} \tanh z/L, \tanh ^2  z/L \right\}
\ea
where  $L$ is a characteristic width of the layer, ${\bf P} $ is a Poynting flux.
Thus, there is a Poynting flux of electro-magnetic energy along the current layer.
At the different side of the current layer $P_y$ changes sing, while $P_x$ 
remains the same.
(Fig. \ref{shear-1}).
The layer is sheared both in $x$ and $y$ direction. On the different sides of the
current layer the plasma is streaming in opposite direction along the $x$, with a
four velocity at infinity reaching $u_x = \pm U_0/\sqrt{1+U_0^2}$, while
along the $y$ direction the plasma is streaming in the same
direction on both sides of the layer, reaching at infinity 
$u_y = - U_0^2/\sqrt{1+U_0^2}$.

Another possible type of layer is a sheared and rotating magnetic field, 
(Blandford, private communication)
\ba &&
\B = B_0  \left\{ \tanh z/L, \sqrt{2} \sech z/L ,0 \right\}
\nn &&
\E = B_0 \left\{ 0,0, \sech z/L\right\}
\nn &&
\J = { B_0 \over L} 
  \left\{  \sqrt{2}  \sech z/L \tanh z/L,  \sech^2 z/L,0 \right\}
\nn &&
\rho_e = - {B_0 \over L} \sech z/L \tanh z/L 
\nn &&
{\bf V} =
\left\{  - {  \sqrt{2}  \over 1+ \cosh^2  z/L}, { \sinh  z/L \over 1+ \cosh^2  z/L},0\right\}
\nn &&
 v^2 = { 1 \over 1+ \cosh^2  z/L}, \,
\gamma^2 = 1+ \sech^2 z/L
\nn &&
{\bf P}= 
\left\{ - \sqrt{2} \sech^2 z/L, \sech z/L \tanh z/L,0 \right\}
\ea
(Fig. \ref{shear-2}).

\item 
{\it Magnetic rotation discontinuity $\E =0$}.
In this case we find $B^2 = B_0^2$, so that magnetic field rotates
keeping its absolute value constant (Fig. \ref{tear-sheet}). 
This case will be our primary interest.
\end{enumerate}

Finally we note, that since ideal electro-dynamics supports only two types
of waves (luminal fast modes and subluminal \Alfven modes), the current sheets
discussed above  are nothing else but  \Alfven waves  considered
in some preferred frame of reference (Komissarov 2002). Then the
instability considered below are related to the instabilities of 1-D  dissipative
 \Alfven waves
(we thank S. Komissarov for pointing this out).

\subsection{Resistive Force-free  Current Sheets}

In a strongly resistive media one may seek stationary solutions of the 
 equation of resistive electro-dynamics  (\ref{Max}, \ref{ffOhm}),
 assuming that resistive terms
are important everywhere (in a way similar to important works of Low (1973) for 
non-relativistic plasma).
Assuming a sheared magnetic field field configuration with 
$E_z=0$ we find 
\ba &&
B_y' = -  {1\over   \eta} { B_x (\E \cdot \B) \over 
\sqrt{ B^2 - (E_y B_x -B_y E_x)^2}}
\nn &&
B_x'=  {1\over  \eta} { B_y (\E \cdot \B) \over
\sqrt{ B^2 - (E_y B_x -B_y E_x)^2} }
\ea
From which it follows that
$B^2 = {\rm const}$.
Assuming $B_x = \cos \phi(z), B_y = \sin \phi(z)$ and 
 setting  without loss of generality
$E_y=0$
we find equation for $\phi$:
\be
\partial_z \phi = - {1\over   \eta} { E_x \cos \phi \over 
\sqrt{ 1- E_x^2 \sin^2 \phi}}
\ee
Which 
can be integrated to give an implicit dependent $\phi(z)$:
\ba &&
{E_x z \over B_0   \eta} =
 - {  E_x  \over B_0} {\rm arcsin} ( {  E_x  \over B_0} \sin \phi) - 
{\sqrt{ 1-  (E_x/B_0)^2}  \over 2} 
\ln {\cal A}
\nn &&
{\cal A}=
\left( {1 \over \cos^2 \phi}
\left( 1+ \left(1- 2 \left({ E_x \over B_0}\right)^2 \right) \sin^2 \phi +
2 \sqrt{ 1-  \left({ E_x \over B_0}\right)^2}  \sin \phi 
\sqrt{ 1-   \left({ E_x \over B_0}\right)^2 \sin^2 \phi } \right)
\right)
\label{Low}
\ea
It describes a inflow of plasma into resistive current layer.
For small $(E_x/B_0) \rightarrow 0$ this gives
\be
\phi =- \tanh {E_x z \over B_0   \eta}
\label{Low1}
\ee
which corresponds to the Low (1973) solution
\ba &&
B_x = B_0 \sech { E_x  \over B_0} {z \over  \eta} 
\nn && 
B_y = - B_0 \tanh { E_x  \over B_0} {z \over  \eta}
\nn &&
v_z = - {E_x \over B_0} \tanh { E_x  \over B_0} {z \over  \eta}
\ea

Application of this solution to astrophysical plasmas suffers from the same
problem as the original Low's solution: the diffusion time scales are much longer than
the time scales of interest.  We suspect (though did not show it) that
the relativistic  solutions (\ref{Low}) is unstable to small
perturbations
which would lead to electric current singularity.
This brings us to the main part of the paper - development of resistive 
 current sheets  in a relativistic force-free plasma.

\section{Tearing mode stability of rotational discontinuities}

A standard method in describing the evolution of tearing mode
is similar to the boundary layer problem 
\footnote{Kinetic approach to tearing mode is described
in in Galeev (1984). It was generalized to relativistic particle
dynamics by Zeleny \& Krasnoslskih (1979).}.
 It 
involves a separation of a current layer into a "bulk", where derivatives
are small and resistivity is not important,  and a   narrow "boundary layer"
where derivatives and resistivity may be large. 
Two different approximations are done in each layer -
 ideal and weakly varying plasma in the bulk
and a narrow resistive  sublayer
(Fig. \ref{tear-sheet}). Two solutions should be matched continuously.


In this section we investigate a resistive stability of a relativistic current layer.
We assume that initially
  the electric field is zero, so that the magnetic field 
while remaining constant in magnitude 
 rotates over some angle (Section \ref{ideal}),
 which we chose to be $\pi$ radians
 for simplicity.

Assume that a current layer has a 
 width $L$. 
Since $B^2=const$, a possible form of the unperturbed magnetic field is 
\ba &&
B_{0,x}= B_0 { z \over L}
\nn &&
B_{0,y} = B_0 \sqrt{ 1- \left( {z \over L} \right)^2}
\ea
Next we investigate stability of such current layer to small perturbations.

\subsection{Stability of resistive  force-free current sheet}

In this main section of the paper we investigate the stability properties
of a planar resistive current sheet.  We assume that in the bulks of the plasma 
resistivity is small, so that initial configuration is described by an ideal
current sheet in which  magnetic field remains planar ($B_z=0$) and
  rotates over some angle   $B^2 = B_0^2$, while electric field is zero $ E=0$.
Formally we should start with resistive solution (\ref{Low})
with has non-zero 
velocity of plasma into the current layer, but for large scale current
that resistive inflow velocity is negligible.

 Consider small fluctuations of the electro-magnetic fields and current
\ba &&
\B = \B_0 +  {\bf b}
\nn &&
\E =  {\bf e}
\nn &&
\J={\bf J}_0 + {\bf j}
\ea
Assume that the perturbations vary as $\exp\{i (\om t - k_x x -k_y y)\}$.
Then eqns (\ref{Max}),  (\ref{Ohm-1}) and (\ref{Ohm-2}) give
\ba && 
b_z \om - (\k \times {\bf e})_z =0
\nn && 
i (\k \cdot {\bf b}) - \partial_z (\k \times {\bf e})_z =0
\nn &&
e_z k^2 + \om  (\k \times {\bf b})_z \om - i  \partial_z  (\k \cdot {\bf e}) =0
\nn &&
(\k \cdot {\bf j}) + i \om (\k \cdot {\bf e}) + \partial_z  (\k \times {\bf b})_z=0
\nn &&
b_z k^2 + i (\k \times {\bf j})_z -
 \om (\k \times {\bf e})_z - i \partial_z (\k \cdot {\bf b})=0  
\nn &&
j_z + i( (\k \times {\bf b})_z + \om e_z)=0
\nn &&
(\B_0 \cdot {\bf e}) -  \eta
( (\B_0 \cdot {\bf j}) - (\B_0 \cdot {\bf b}) H )=0
\nn && 
j_z + b_z H =0
\ea
where $H= (B_x \partial_z B_y - B_y  \partial_z B_x)/B_0^2$ is a normalized current
vorticity of the initial state and $\k =\{k_x,k_y,0\}$.

We can readily eliminate current,   $ (\k \times {\bf j})_z, \,
(\k \cdot {\bf j}), \, j_z $, as  well as 
z-components of electric and magnetic field fluctuations $e_z$ and $b_z$.
After some rearrangements the equations involving fluctuations of
$x$-  and $y-$ components of electric and magnetic fields become
\ba && 
i (\k \cdot {\bf b}) - \partial_z (\k \times {\bf e})_z =0
\nn &&
i \om 
(\k \times {\bf b})_z \left(1 -{ \om^2 \over k^2} \right)
- H \om  (\k \times {\bf e})_z  - { \om^2 \over k^2} \partial_z (\k \cdot {\bf e}) =0
\nn &&
(\B_0 \cdot {\bf e}) -   \eta
\left( i  (\k \times \B_0)_z (\k \times {\bf e})_z   -
2 \om H (\B_0 \cdot {\bf b}) - \om \partial_z (\B_0 \times {\bf b})_z \right) =0
\nn &&
i  (\k \cdot \B_0) (\k \times {\bf e})_z - i \om^2 (\B_0 \times {\bf e})_z + 
\partial_z (\B_0 \cdot {\bf b})=0
\ea

From which we can eliminate magnetic fields
\ba &&
(\B_0 \cdot {\bf b}) =- i \left( 
{ (\k \cdot \B_0)  \partial_z (\k \times {\bf e})_z \over k^2 
\om} 
 + { (\k \times \B_0)_z \left( H k^2  (\k \times {\bf e})_z  + \om^2  
\partial_z (\k \cdot {\bf e})  \right) \over
k^2 \om (k^2 - \om^2)} \right)
\nn &&
( \B_0 \times {\bf b})_z =
i \left( { (\k \times \B_0)_z \partial_z  (\k \times {\bf e})_z \over k^2 \om} -
{(\k \cdot \B_0) \left( H  (\k \times {\bf e})_z  + \om^2/k^2 
 \partial_z  (\k \cdot {\bf e})
\right)  \over \om (k^2 -\om^2)} \right)
\label{A}
\ea

This, with corresponding equations for electric fields
\ba && 
(\B_0 \cdot {\bf e}) -    \eta
\left( i  (\k \times \B_0)_z (\k \times {\bf e})_z - 2 H \om (\B_0 \cdot {\bf b})
- i  \om^2  (\B_0 \cdot {\bf e}) - \om \partial_z ( \B_0 \times {\bf b})_z\right) =0
\nn && 
i  (\k \cdot \B_0) (\k \times {\bf e})_z - i \om^2 (\B_0 \times {\bf e})_z +
\partial_z (\B_0 \cdot {\bf b})=0
\label{B}
\ea
is where we stop the general approach.

\subsection{Ideal current layer}

In the bulk of the plasma we can neglect resistivity so that the dynamics
of the current layer is governed by equations (\ref{Max},\ref{FF}). 
In this section
we investigate stability of an ideal force-free current sheet to  small 
 perturbation 
(see also Gruzinov 1999).

For ideal plasma $\eta =0$, we find 
\be
\partial_z \left(F \partial_z {e_y  \over B_x \om } \right)+
{ e_y \over   B_x \om } F (\om^2 -k^2)=0
\label{stabideal}
\ee
Or, returning to the definition of $b_z$
\be
\partial_z \left(F \partial_z {b_z  \over (\B_0 \cdot \k) } \right)+
{ b_z \over  (\B_0 \cdot \k) } F (\om^2 -k^2)=0
\label{stabideal1}
\ee
where $F=  (\B_0 \cdot \k) ^2 - \om^2 B_0^2$.
Equation (\ref{stabideal1}) describes stability properties of
an ideal relativistic force-free current sheet.
 A quick examination shows that similar to the non-relativistic case  points
$(\B_0 \cdot \k)=0$ is  a special point of the equation.
Fluctuation of the magnetic field $b_z$ diverge at these surfaces 
creating current sheets.

Consider fluctuations near the surfaces $(\B_0 \cdot \k)=0$.
Assume that $B_x $ becomes zero at $z=0$; then,
neglecting variations  in $y$ direction, $k_y \rightarrow 0$ and 
introducing displacement $\xi = b_z /B_x k_x$
we find
\be
 \left(B_x^2 k_x^2 -B_0^2 \om^2 \right) 
\left( \xi ( \om^2 -k_x^2 ) + \partial_z^2 \xi \right) 
+ 2 B_x k_x^2  \partial_z \xi \partial_z B_x
\label{Z0}
\ee
which can be further simplified 
\be
(\hat{k}_x^2 - \hat{\om}^2 ) \left( \hat{B}_x^2 \hat{k}_x^2 - \hat{\om}^2 \right)
\xi(Z) - \partial_Z^2 \xi(Z)
\label{Z}
\ee
where 
\ba &&
\hat{k}_x = k_x L 
\nn &&
\hat{\om}=\om L
\nn &&
\hat{B}_x={B_x \over B_0}
\nn &&
\hat{z}= { z \over L} 
\nn &&
Z=  \int {1\over \hat{k}_x^2  \hat{B}_x^2 - \hat{\om}^2}  d \hat{z}
\ea

For example, for a linear $\hat{B}_x = \hat{z}$, we find
$Z=- {\rm arctan} (\hat{k}_x \hat{z} /\hat{\om})/ (\hat{k}_x \hat{\om})$. Then
eq. (\ref{Z}) gives
\ba
\hat{\om}^4( \hat{k}_x^2 - \hat{\om}^2 ) \sech^2 ( \hat{k}_x \hat{\om} Z) \xi - 
\partial_Z^2 \xi(Z) =0
\label{Z1}
\ea
which has a form of a non-linear oscillator.

General solution 
of Eq. (\ref{Z1}) is quite complicated. Its dependence on $\hat{\om}$
shows that variations of the field have electro-magnetic structure
(as oppose to purely magnetostatic variations in the non-relativistic case).
We  expect that typical growth rates will be small $\hat{\om} \ll  \hat{k}_x$,
so
that we can neglect the electro-magnetic corrections by setting
$\hat{\om} \rightarrow 0$. This will have to be checked {\it a posteriori}.
This limit corresponds to non-relativistic approximation, so that the
relations (\ref{P1}-\ref{P2})  coincide with the familiar relations for the
non-relativistic tearing mode.

In the limit $\hat{\om} \rightarrow 0$ eq. (\ref{Z0}) gives for
$\hat{B}_x = \hat{z}$ when $|\hat{z} | <1$ and $\hat{B}_x =1$ for $|\hat{z} | >1$
\be
\xi = 
\left\{ \begin{array}{cc} 
C_1 { \cosh \hat{k}_x \hat{z} \over \hat{z}} + 
C_2 { \sinh \hat{k}_x \hat{z} \over \hat{z}} & 
\mbox{ if $ |\hat{z} | <1$ } \\
C_0 \exp\{- \hat{k}_x |\hat{z}|\} & \mbox{ if $|\hat{z} | >1$}
\end {array} \right.
\label{P1}
\ee
where we  have chosen a solution which is decaying as $ |\hat{z} |\rightarrow \infty$.
Matching the two solutions at $|\hat{z}|=1$ gives
\ba &&
C_1 = {C_0 \over 2 \hat{k}_x} \left( 2 \hat{k}_x -1 + \exp\{-2 \hat{k}_x \} \right)
\nn && 
C_2 = {C_0 \over 2 \hat{k}_x} \left(1-  2 \hat{k}_x + \exp\{-2 \hat{k}_x \} \right)
\ea
 Given a solution for $\xi$ we can find magnetic field fluctuation
\be
{ b_z  \over B_0} = 
\hat{B}_x \hat{k}_x \xi =
 \hat{k}_x \left(C_1  \cosh \hat{k}_x \hat{z}  + C_2 \sinh \hat{k}_x \hat{z} \right)
\ee
Magnetic field is continuous at $\hat{z} =0$ but its
derivative experiences a jump 
\be
\Delta \equiv \left[ { \partial_{\hat{z}}  b_z \over b_z } \right]
={2 C_2 \hat{k}_x \over C_1} \approx
 {2 \over \hat{k}_x}  \mbox {,  for $\hat{k}_x \ll 1$}
\label{P2}
\ee

This means that there is  current sheet forming at $\hat{z}=0$. 

\subsection{Structure of resistive sublayer}

To study the structure of the resistive sublayer we make several approximations
for equations  (\ref{A}-\ref{B}).
First, we neglect variations along the $y$-axis, $k_y \rightarrow 0$. 
We find then
\ba &&
e_x B_x + e_y B_y  +    { \eta \over \Gamma}  
\left(
e_y \left( B_y(H^2+k_x^2) + B_x  \partial_z H \right)
+ 2 B_x H  \partial_z e_y  - B_y  \partial_z^2 e_y +
\right. 
\nn &&
\left. 
{ \Gamma^2 \over k_x^2 + \Gamma^2}
\left( (B_x e_x + B_y e_y) (k_x^2 + \Gamma^2) + ( B_y H^2+ B_x \partial_z H)  e_y +
H ( B_y \partial_z e_x + B_x \partial_z e_y) +B_x \partial_z^2 e_x \right)
\right)=0
\nn &&
e_y \left( B_x (k_x^2 -H^2) - B_y \partial_z H \right) - B_x \partial_z^2 e_y
+
\nn &&
{\Gamma ^2 \over k_x^2 + \Gamma^2 }
\left( (e_y B_x - B_y e_x) (k_x^2 + \Gamma^2)+
e_y (B_x^2 H^2 +B_y \partial_z H) +
H ( B_x \partial_z e_x + B_y \partial_z e_y ) + B_y \partial_z^2 e_x \right)
=0
\label{ZX}
\ea
where $\Gamma = - i \om$.

We expect that the width of the resistive layer, $\sim \epsilon L$ is much smaller than the 
width of the current layer $L$, $\epsilon \ll 1$. We then 
expand system (\ref{ZX}) near $z \sim \epsilon $ by introducing 
 $\zeta= \epsilon  z /L$. 
  
For $ \epsilon \ll 1 $ 
we find
\ba && 
e_y -e_x \zeta \epsilon - \hat{\eta} { \partial_\zeta^2 e_y \over \epsilon^2}=0
\nn &&
\zeta \left(
e_y \hat{k}_x^2 \epsilon^2 - {\partial_\zeta^2 e_y } \right)
+ {\hat{\Gamma}^2 \over \hat{k}_x^2 } 
\left( { \partial_z^2 e_x \over \epsilon} -
{\partial_\zeta ( \zeta \partial_\zeta  e_y ) \over \zeta} 
\right)=0
\label{R}
\ea
where $\hat{\eta}=   
 \eta /(\Gamma L^2)$, $\hat{\Gamma}= \Gamma L$, $\hat{k}_x = k_x L $
($\hat{\Gamma}= \Gamma L/c$ in dimensional units)
and we used $ B_x \sim B_0 z, B_y \sim B_0, \partial_z H \sim - z/L^2, H \sim -1/L$.
We also assumed $\hat{\eta},\, \hat{\Gamma}\ll 1$ and neglected
terms of higher orders in parameters $\hat{k}_x, \hat{\eta}, \hat{\Gamma}^2$.

Eliminating $e_x$ we find a fourth order equation for $e_y$
(in which we can also neglect $\zeta \epsilon \ll 1 $ and
$ \hat{\Gamma} \ll \hat{k}_x$):
\be 
\zeta \hat{\Gamma}^2 \hat{\eta} \left(
\zeta \partial_\zeta^4 e_y - 
2  \partial_\zeta^3 e_y \right) 
+ 
\left( 2    \hat{\Gamma}^2 \hat{\eta} - 
\zeta ^2  \epsilon^2 ( \hat{\Gamma}^2  + \zeta ^2   \hat{k}_x^2  \epsilon^2)
 \right) \partial_\zeta^2 e_y
+ 4 \zeta  \hat{\Gamma}^2  \epsilon^2   \partial_\zeta e_y
- \epsilon^2 \left( 2 \hat{\Gamma}^2 -  \hat{k}_x^4 \zeta ^4  \epsilon^4 \right)
=0
\label{Big}
\ee
The typical scales appearing in  eq. (\ref{Big}) are
\ba &&
\delta_1  = { \sqrt{ \hat{\eta}}  \over \epsilon} 
= \sqrt{ { \eta \over  \epsilon \Gamma L^2 }}
\nn && 
\delta_2 = {  \sqrt{  \hat{\Gamma} } \hat{\eta}^{1/4} \over \sqrt{ \hat{k}_x} \epsilon }=  \sqrt{ { \hat{\Gamma} \delta_1 \over \epsilon \hat{k}_x} } =
{ ( \eta \Gamma  /L^3 c)^{1/4} \over \sqrt{ k_x L} }  
\nn && 
\delta_3= {  \sqrt{  \hat{\Gamma} } \over \epsilon \hat{k}_x } =
 {  \sqrt{  \Gamma /(L c)  } \over \epsilon  k_x L  }
\ea
$\delta_1$ is  a resistive skin depth  (the distance that the field diffuses
in time $\Gamma^{-1}$), $\delta_2$ is a width of tearing sub-layer,
$\delta_3 \ll \delta_1$ does not seem to have a physical meaning.

For very small $\zeta \rightarrow 0$ eq (\ref{Big}) becomes
\be
\partial_\zeta^2 e_y - { \epsilon^2 \over \hat{\eta}} e_y =0
\ee
which has a solution
\be
e_y \sim \cosh { \zeta  \epsilon \over \sqrt{\hat{\eta} }}
\ee
where we chose $ e_y' (\zeta=0)=0$.
This solution is valid for $\zeta \ll \delta_2$. 
For the order-of-magnitude estimates we can assume that this solution is valid
until $\zeta \sim  \delta_2$. 
Setting $\delta_2=1$ we find the  width of the resistive layer:
\be 
\delta/L = \epsilon =
 \sqrt{ \hat{\Gamma} \over \hat{k}_x }  \hat{\eta}  =
 \sqrt{ { \hat{\Gamma} \delta_1 \over \hat{k}_x} }
\label{ep1}
\ee
(compare with Woods (1987), eq. 7.116).

Equating then $\Delta = e_y'/e_y$ taken at $\zeta=  1 $ 
to external solution we find
\be 
\Delta ={\epsilon ^2   \over \hat{\eta}} = { \epsilon  \over \hat{k}_x}
\ee
From which we find 
\be
\epsilon  \sim  {  \hat{\eta} \over \hat{k}_x} 
\label{ep2}
\ee
The growth rate  then  follows from eqns (\ref{ep1}) and (\ref{ep2})
\be
\Gamma \sim \left( { \eta ^3 c^2  \over k_x^2 L^{10}  } \right)^{1/5} = 
{ 1 \over k_x^2 L^2} 
{ 1 \over \tau_R^{3/5} \tau_A^{2/5}} 
\ee
The growth rate increases with $1/\hat{k}_x$. The maximum rate may be
found from the condition  that  the resistive time scale for a sublayer of width
$\epsilon L $,  is much smaller that the  growth rate:
\be
 \Gamma \leq (k_x L)^{2/3}  \left( { \eta c^2  \over L^4 } \right)  ^{1/3} 
\label{gamma1}
\ee
Which gives 
\be 
k_x L \sim  \left(  { \eta \over L c} \right) ^{1/4}
\ee
and
\be
\Gamma \sim \sqrt{ \eta c \over L^3} \sim {1 \over \sqrt{ \tau_R \tau_A}}
\label{gamma}
\ee
This estimate formally 
 coincides with the case of non-relativistic non-force-free
plasma. 

\section{Application to magnetars}

One of main shortcomings
of the current approach is that resistivity $\eta$ was not 
calculated from the particle
 kinetics, but was introduced as a macroscopic property
of plasma - a common approach in continuous mechanics. Resistivity in 
tenuous astrophysical  plasmas is due to
 collective processes and not binary collision.
It has to be excited by plasma currents and thus is likely to be a (non-linear)
function of a current itself. This requires a kinetic treatment of plasma-wave 
interaction as well as a
correct account of  non-linear (or quasi-linear) feedback
of plasma turbulence on the particle. 
Excitation of plasma turbulence by currents in force-free fields
 is confirmed by a number  of  electro-magnetic particle-in-cell simulations
(\eg Sakai \etal 2001). Qualitatively,  
when the drift velocity of a current exceeds the plasma thermal velocity, 
strong plasma turbulence is excited.

In spite of complicated microphysics which determines  the 
resistivity
we can make qualitative upper estimates on the value of resistivity $\eta$.
At the early stages of the development of instability the 
  plasma remains force-free, until a large amount of magnetic energy
 has been dissipated. Since under force-free
 assumption the particles are bound to move
only along the field lines, 
this limits considerably a number of possible resonant 
wave-particle interactions that can lead to development of plasma turbulence.
The two remaining options 
are the Langmuir turbulence, which  in relativistic  plasma
develops on a typical scale of electron skip depth, $\delta_e \sim c/\om_{p,e}$,
and ion sound  turbulence, which  in relativistic plasma develops on an ion skip depth
$\delta_i \sim c/\om_{p,i}$ ($\om_{p,e}$ and $\om_{p,i}$ are the electron and ion plasma
plasma frequencies). They are different by a square root of the ratio of electron
to ion masses $\delta_e/\delta_i \sim (m_e /m_i)^{1/2}$.
  Qualitatively, a fully developed turbulence
with a typical velocity $c$ and  typical scale $\delta$ would produce a resistivity
\be 
\eta \sim c \delta
\ee
Using this estimate in 
(\ref{gamma}) we find
\be
\Gamma \sim {c \over L} 
\left( { \delta \over L} \right)^{1/2}
\ee
Note, that 
 the growth rate of the tearing mode is proportional to $ (m_e /m_i)^{1/4}$,
 which
is only a factor of few.

In current-carrying magnetospheres of  magnetars the plasma frequency
is (this estimate comes from he requirement that toroidal magnetic field
created by currents does not exceed poloidal,  Thompson \etal 2002)
\be
\om_p \sim \sqrt{ \om_B r/c} \sim 2 \times 10^9\, {\rm rad/sec}
\ee
for $B \sim 5 \times 10^{14}$ G and $r\sim 10^{7}$ cm.

Then
\be
 \delta \sim 10\, {\rm cm}, \,
\eta \sim 3 \times 10^{11} {\rm cm^2 /sec}
\ee
Resistive  and \Alfven time scales are  then
\ba &&
\tau_R \sim L^2 / \eta =  3 \, L_{10^6}^2\, {\rm sec} 
\nn &&
\tau_A \sim L/c =  3 \times 10^{-5} \, L_{10^6}\, {\rm sec} 
\ea
where 
$L_{10^6}= L /10^6 {\rm cm} $.

The tearing mode growth time is then
\ba
\tau_t \sim
{1 \over \Gamma } \sim  10^{-2} \, {\rm sec}\,  L_{10^6}^{3/2}
\label{tau_t}
\ea
For a given $L$  this is  the lower estimate on $\tau_t$ since we used
an upper estimate on $\eta$.

The growth  time (\ref{tau_t}) is of the order of  the observed
 rise time of the SGR X-ray
flares, $\leq 10$  msec. For a given current sheet width $L$ eq.
 (\ref{tau_t}) gives a lower estimate and also in
Since in 
the observed bursts  the rise time
is limited by the intensity of the burst -
weaker bursts are expected to have shorter rise times (G\"og\"us 2002) - smaller
burst should come from smaller current sheets.

\section{Discussion}
\label{Discussion}

In this paper we have first formulated equations of resistive force-free
electro-dynamics, which describes strongly magnetized  dissipative 
relativistic  plasmas. This is an important step towards understanding 
of plasma dynamics in extreme conditions (when energy density is dominated
by magnetic field energy density) and may  serve as a guiding rule for 
numerical investigations
  of a wide variety of
 electro-magnetically dominated astrophysical plasmas (since most
numerical schemes are necessarily dissipative).

Next 
we have analyzed  development of 
unsteady reconnection in a relativistic force-free
plasma. 
We find that under assumptions  of a  force-free resistive plasma the 
development of  a tearing mode proceeds qualitatively in a way similar to the
non-relativistic case, when inertia of matter is not  important. 
This is a bit surprising result, given that we solve a quite different system
of equations.  Several factors may explain this.
First, the special relativistic corrections
( $1/\sqrt{1- (\E \times \B)^2/B^2}$ term in (\ref{ffOhm})
 did not enter our linearized
 analysis 
since we were considering pure magnetic equilibrium,
 without an equilibrium drift of
particles. Thus, the 
 solution for the bulk of the current layer virtually coincides with the
 non-relativistic case, since even in the  non-relativistic case 
the motion in the bulk of the current layer is assumed to be inertia- and
resistance-free, thus obeying ideal force-free equations.
Secondly, 
in relativistic regime the important 
quantity that controls the dynamical response
of a plasma to  a given  force is not inertia but energy density. In force-free
plasmas the energy density is completely composed of magnetic field (one may say
that magnetic field    gives
an effective inertia to plasma).

We find that  the typical  rise time of SGR flares, $\sim 10 $ msec may be related
to the development of unsteady reconnection  via tearing mode in the strongly magnetized 
relativistic plasma of magnetar magnetospheres. To obtain this estimate we have assumed that
the plasma resistivity is provided by relativistic  Langmuir turbulence with a typical 
spatial scale of the order of the skin depth, $\sim c/\om_p$,  that the corresponding
 plasma
density is found using the Thompson \etal (2002)
 model of current-carrying magnetosphere, 
and that  the resistivity is given by the fully-developed
relativistic  Langmuir 
turbulence, $\eta \sim c^2 \om_p$.

The 
relation between the rise time of a flare and the total energy output 
cannot be simply predicted in this model. On one hand, 
growth rate is independent of the magnetic field and the value of the current
in the layer. On the other hand, we expect that the resistivity $\eta$ 
is  anomalous resistivity excited by currents and thus may correlate with
current strength and the amount of energy stored in non-potential
magnetic fields. 
In addition,
very long burst with multiple components should result from numerous
avalanche-type
reconnection events, as reconnection at one point may triggers
reconnection at other points (\eg as in the SOC model of  Li \& Hamilton 1991).

Development of tearing mode is likely to be 
 accompanied by acceleration of particles
and production of high energy emission similar to the well studied
 acceleration in the Earth
magnetotail (\eg Coroniti and Kennel 1979).
Qualitatively, acceleration may be done by  electric fields
directed along magnetic fields, or near magnetic null surfaces.
 At later stages of reconnection
various  mechanisms of particle acceleration (DC electric fields, stochastic
acceleration,
shock acceleration) may be operational (\eg Aschwanden  2002).

An alternative possibility for production of flares on magnetars  is that flares
 may result from a sudden change (unwinding) in the
internal magnetic field.  In this case, a twist is implanted into
the magnetosphere. 
 A large-scale displacement of the crust probably
requires the formation of a propagating fracture, close to which
the magnetic field is
 strongly sheared (Thompson \& Duncan 1995, 2001; Woods et al. 2001a).
For this scenario it is important that the critical  crustal shear stress 
be relatively large, $\sim 0.1$. It is hard to estimate the critical stress
for the neutron star crusts (critical stress cannot be calculated from first
principles). Indirect evidence based on possible measurement of free
precession (Cutler \etal 2002)  show that critical stress is small,  $\sim 10^{-5}$, 
favoring  the plastic creep possibility.

  Below 
we  summarize  the evidence that  point to the magnetospheric origin of flares
(see also Thompson \etal 2002).
\begin{itemize}
\item   The pulse profile of SGR 1900$+$14 changed dramatically following
the August 27
giant flare, simplifying to a single sinusoidal pulse from 4-5 sub-pulses
(Woods et al. 2001a). 
In the reconnection  model the post-flare magnetosphere 
 is expected to have  a simpler  structure, as the pre-flare network of currents
has been largely dissipated.
\item  The persistent spectrum of SGR 1900$+$14 softened measurably
after the 27 August giant flare: the
best-fit spectral index (pure power law) softened from
-$1.89\pm 0.06$ to -$2.20\pm 0.05$  (Woods et al. 1999a).
Since the spectral index is a measure of the current strength in the 
magnetosphere (Thompson \etal 2002), this points to weaker currents in the 
post-flare state, consistent with dissipation of currents during a flare.
\item The typical rise-time of bursts, $\sim 0.01$ sec is consistent with the 
time-scale for the development of force-free tearing mode in the magnetosphere. 
\item  SGR bursts come at random phases in the pulse profile Palmer 
(2000) - this is naturally
explained if (even  only one!) 
 emission cite is located high in the magnetosphere, so that we see all
the bursts 
 (if the bursts were associated with a particular active region
on the surface of the neutron star,
 one would expect a correlation with a phase);
\item  pulsed fraction increases in the tails of the strong bursts, keeping
the  pulse profile similar  to the persistent emission (Woods \etal  2002) - 
this is easier explained if the energy release processes occurring  high in 
the magnetosphere after the giant burst 
 are connected to the same hot spot on the surface of the neutron star as the field
which are active during the quiescent phase.
\item   smaller fluency SGR events, have
harder spectra than the more intense ones (G\"og\"us \etal 2002) (this is 
 also true for the spikes of multi-structured bursts);
 this is consistent with
short events being due to reconnection,  while longer events have a
large contribution from the surface, heated by the precipitating particles.
\end{itemize}

We  will also  make several qualitative suggestions regarding the
possible relation between AXPs and SGRs in a frame-work of magnetospheric
model for energy dissipation. The prime question is why AXPs and SGRs
look so different (AXPs emit mostly persistent emission, while SGRs are actively 
bursting) in spite of  similar  physical conditions.
First, following the Parker's paradigm  for  Solar flares, it is possible that the 
persistent  emission
of magnetars, as well as bursts, is powered by small scale reconnection events.
The quasi-stable profiles are then due to multiple  scattering of radiation
 in the magnetosphere at large
radii $r \sim 10 r_{S}$ (Thompson \etal 2002).
 At these radii the field is dominated by lower multipoles, 
which dissipate on longer time scales than the higher multipoles.
Secondly,  AXPs have stronger than SGRs measured magnetic fields and thus can support 
larger  non-potential 
magnetic fields and  larger currents
flowing  through magnetosphere. Larger currents create higher level of turbulence which
contributes to larger resistivity.  In a medium with larger resistivity the 
intermitency is less pronounced, so that dissipation of currents in AXPs is dominated
by small scale reconnection events, while in SGRs bursts on average contribute as much  
energy as persistent emission.

There is a number of relevant problems to be solved. 
Primarily, there is a need for simulations
 to confirm the growth rates and to test the
non-linear stages of the instability. This requires development of 
electro-magnetic codes. We are aware of the
 ongoing development of ideal electro-magnetic  codes, but none
have been completed so far (MacFadyen, private communication; 
Spitkovsky, private communication).  Several exact 
equilibrium solutions presented in this
work 
 offer good tests for numerical schemes. 
Secondly, resistive 
instabilities of relativistic cylindrical plasmas need to be studied as well
in connection with the possible role of reconnection in AGN and pulsar jets.
One expects that in  a  cylindrical geometry the  growth of resistive 
instabilities will be higher than in the planar case. 
Non-linear development of resistive instabilities is also of interest, but 
given a considerable complications involved in calculating non-linear stages
of instabilities and obvious difficulties in relating
the results to observations, it is not considered a promising approach
(at least analytically)  at the moment.

\acknowledgments
I would like to thank Eric Blackman, Roger Blandford, Sergei Komissarov,
 Anatoly Spitkovsky,  Vladimir Pariev, Dmitry Uzdensky 
for their interest in this work.

\end {document}